# Alignment of carbon nanotube additives for enhancing the magnesium diboride superconductors' performance**

Shi Xue Dou*, Waikong Yeoh, Olga Shcherbakova, David Wexler, Ying Li, Zhong M. Ren, Paul Munroe, Sookien Chen, Kaisin. Tan, Bartek A. Glowacki and Judith L. MacManus-Driscoll

The rapid progress on $MgB_2$ superconductor since its discovery[1] has made this material a strong competitor to low and high temperature superconductors (HTS) for applications with a great potential to catch the niche market such as in magnetic resonant imaging (MRI). Thanks to the lack of weak links and the two-gap superconductivity of $MgB_2$[2,3] a number of additives have been successfully used to enhance the critical current density, $J_c$ and the upper critical field, $H_{c2}$.[4-12] Carbon nanotubes (CNTs) have unusually electrical, mechanical and thermal properties[13-16] and hence is an ideal component to fabricate composites for improving their performance. To take advantages of the extraordinary properties of CNTs it is important to align CNTs in the composites. Here we report a method of alignment of CNTs in the CNT/$MgB_2$ superconductor composite wires through a readily scalable drawing technique. The aligned CNT doped $MgB_2$ wires show an enhancement in magnetic $J_c(H)$ by more than an order of magnitude in high magnetic fields, compared to the undoped ones. The CNTs have also significantly enhanced the heat transfer and dissipation. CNTs have been used mainly in structural materials, but here the advantage of their use in functional composites is shown and this has wider ramifications for other functional materials.

---------------------------------
[*]     Prof. S.X. Dou, W.K. Yeoh, O. Shcherbakova and Dr. D. Wexler
        Institute for Superconducting and Electronic Materials, University of Wollongong, Northfields Ave. Wollongong, NSW 2522, Australia
        E-mail: shi_dou@uow.edu.au
        Dr. Y. Li and Z.M. Ren
        Department of Mater. Sci. and Eng., Shanghai University, 149 Yanchang Rd. Shanghai 200072 P.R. China

Prof. P. Munroe
Electron Microscope Unit, University of New South Wales, Sydney, NSW 2000, Australia
S.K. Chen, K.S. Tan, B.A. Dr. Glowacki and Dr. J.L. MacManus-Driscoll
Department of Materials Science and Metallurgy, University of Cambridge, Pembroke St. Cambridge CB2 3QZ, UK

[**] The work was supported by the Australian Research Council, Hyper Tech Research Inc, OH, USA, and Alphatech International Ltd, NZ. W.K.Yeoh received an Australia-Asia Award funded by the Australian Government. Z.M.R and Y.L thank the Natural Science Foundation of China for their support under grants (No. 59871026, 50225416, 50234020). K.S.Tan received support from ORS/Cambridge Commonwealth Trust. Funding from EPSRC in the U.K. is also acknowledged

High in-field $J_c$ of $MgB_2$ superconductors is one of the major requirements for all large scale applications. In addition, the $MgB_2$ wires should have good mechanical properties and thermal stabilities. The additives studied so far have been mainly used to improve the $J_c(H)$ performance by introducing effective pinning sites in the $MgB_2$. Little attention has been given to the mechanical and thermal properties of $MgB_2$ wire core. Among these additives studied CNT may have the potential to improve the mechanical and thermal properties of $MgB_2$ wires since CNTs have been used as component for a number of composites to improve their mechanical properties.[17,18] Fosshein et al report an enhanced flux pinning in $Bi_2Sr_2CaCu_2O_{8+x}$ superconductors with embedded CNTs.[19] Yang et al found a significant enhancement in $J_c(H)$ for high temperature superconductor by introducing nanorods as columnar pinning centers in to the composites.[20-22] In a previous work we reported that CNT doping enhanced $J_c$ in magnetic fields of bulk $MgB_2$.[12,23] However, in that case the CNTs are randomly dispersed in $MgB_2$ bulk. Prompted by these previous reports, in this work we intend to develop a technique to align the CNTs in Fe sheathed $MgB_2$ wire. The advantages of CNT doping over other additives can be described as follows. (1) Multiwalled CNTs can carry current densities up to $10^9 – 10^{10}$ A/cm$^2$ (compared to a typical value of $10^5 – 10^6$ A/cm$^2$ for superconductors) and remain stable for extended period of time.[14] The CNT doping can improve the current path and connectivity between grains in $MgB_2$. (2) The thermal conductivity for isolated multiwall CNTs is estimated to be about 3000W/mK [15] and hence could benefit the heat dissipation and thermal stability of the MgB$_2$ wires . (3) Bundled CNTs have very high axial strength and stiffness, approaching that for ideal carbon fiber. [16] If CNTs can be aligned along the MgB$_2$ wire longitudinal axis CNTs would improve the mechanical properties of the CNT/MgB$_2$ composite wire. Finally, CNT is typically one-dimensional with high aspect ratio of several orders of magnitude and can act as line pinning sites against the point defect pinning sites of other additives. We demonstrate that a high degree of CNT alignment can be readily achieved by mechanical drawing in the powder in tube (PIT) process. The aligned CNT doped MgB$_2$ wire shows a significant enhancement in flux pinning and heat transfer.

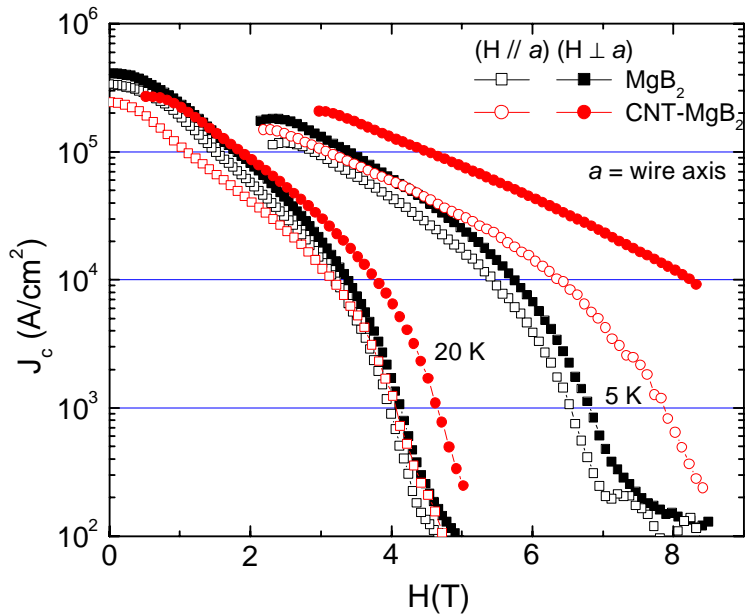

Fig. 1 Magnetic critical current density as a function of applied magnetic field at 5 K and 20 K for the undoped and CNT doped MgB$_2$ wires processed at 800°C for 30 min. All the samples made for magnetic measurement have the same dimension of 0.7mm OD and 2.7mm in length. The measurement field $H$ was applied perpendicular and parallel to the wire axis, $a$, during the measurement of $M$-$H$ loops.

Fig. 1 shows the magnetic $J_c$ calculated from magnetization hysteresis loops with magnetic field, $H$, applied both parallel to and perpendicular to the wire sample axis, $a$, for both undoped and CNT doped MgB$_2$ wires. It is noted that for the undoped MgB$_2$ wire there is a small difference in $J_c(H)$ in relation to the measuring field direction due to the wire drawing process. In contrast, for the CNT doped MgB$_2$ the magnetic $J_c(H)$ shows two distinguished features: a strong anisotropy in relation to the measuring field direction and a clear enhancement of flux pinning in high field region compared to that of the undoped sample. The anisotropy in magnetic $J_c$ at 5 K is stronger than that at 20 K and the extent of the anisotropy in $J_c$ increases with magnetic field $H$, ranging from a factor of 2-3 in low fields up to more one order of magnitude in high fields. The enhancement in $J_c(H)$ performance due to CNT doping is more significant at low temperatures and higher fields. For example, the magnetic $J_c$ in the $H \perp$ direction increases by a factor of 4 at 20 K and 4 T, and 37 at 5 K and 7 T respectively, as a result of CNT doping. What is more striking is that the enhancement in $J_c$ of the CNT doped $MgB_2$ occurs in two directions of applied field H respective to the wire axis. $J_c(H \perp a)$ shows more significant enhancement than $J_c(H//a)$ compared to that of undoped $MgB_2$ wire.

For the reason of anisotropy in the magnetic $J_c$ both XRD and electron diffraction patterns show no evidence for the crystalline orientation of $MgB_2$ grains in the wire core in both longitudinal and transverse direction to the wire axis. Because the formation $MgB_2$ crystal is accomplished through reaction *in situ* process in which there is no driving force for the $MgB_2$ crystalline alignment. Thus, the anisotropy is not due to the $MgB_2$ crystalline alignment. Fig. 2 (a) shows scanning electron micrographs for the CNT doped $MgB_2$ wire. A strong elongated macrostructure along wire axis is clearly evident. In the CNT doped bulk materials the CNT are randomly dispersed in the $MgB_2$ matrix and most CNT are highly entangled as shown in Fig. 2 (b). In contrast, for the powder-in-tube (PIT) process the mixture of Mg and B powder and the CNT additives flows along the wire axis as a result of continuing drawing and reduction processes. The drawing process is repeated by more than 30 times to reduce the iron sheath/Mg + B powder composite from 10 mm diameter to 1 – 1.4 mm diameter. Because of the high aspect ratio of CNT the drawing force drags the entangled CNT to straighten in the longitudinal direction. The degree of CNTs alignment increases with increasing number of drawings. For a single filament $MgB_2$ wire, the filament diameter is relatively large compared to that of CNTs. However, practical conductors are made of multifilament with a diameter of each filament as small as 10 µm which is comparable to the CNTs' diameter (up to 2 µm). Thus, a high degree of CNT alignment can be achieved. Nevertheless, the extent of CNT's alignment is clearly shown from both magnetic $J_c$ measurement and TEM observation even in the single core filament. Fig. 2 (c) shows a TEM image of the embedded, straightened CNTs in the same direction in the $MgB_2$ wire core. The inset shows the high resolution image of the CNT lattice. TEM image for several parallel CNTs in $MgB_2$ matrix. The inset in Fig 2 (d) is the high resolution lattice image of one of CNTs. These results indicate that the anisotropy in $J_c$ originates from the alignment of CNT during mechanical drawing process. The CNT induced anisotropy in $J_c$ may be attributable to two factors: high axial conductivity[14,24] and large aspect ratio of CNTs. The aligned CNTs along the core axis will improve the connectivity between grains by bridging poorly connected regions, which are common problem for the *in situ* reaction process because the density of $MgB_2$ can only reach about 50% of theoretical density in this process. The better in-field $J_c$ performance in the longitudinal direction, when field H is applied perpendicular to the wire axis, suggests that the presence of the CNTs in the longitudinal direction induces strong pinning sites in transverse direction to the longitudinal axis. This comes a surprise as we would expect other way around, that is, the stronger pinning should be in the CNTs' longitudinal direction when field H is applied parallel to the wire axis because flux lines would have strong interaction with CNTs. However, we must consider when the CNTs are aligned in the wire axis direction the total cross section area of CNTs in transverse direction to the wire axis is much smaller than that in the longitudinal direction due to the high aspect ratio of CNTs. Thus, the overall interaction between CNTs and flux lines is stronger when the field H is applied perpendicular to the longitudinal wire axis than parallel to the wire axis, and hence the $J_c$ in longitudinal direction shows better in-field performance than that in transverse direction. As reported previously, the enhancement in $J_c$ by CNT's is attributed to two factors: carbon substitution for B and CNTs as strong pinning centers.[12,23]

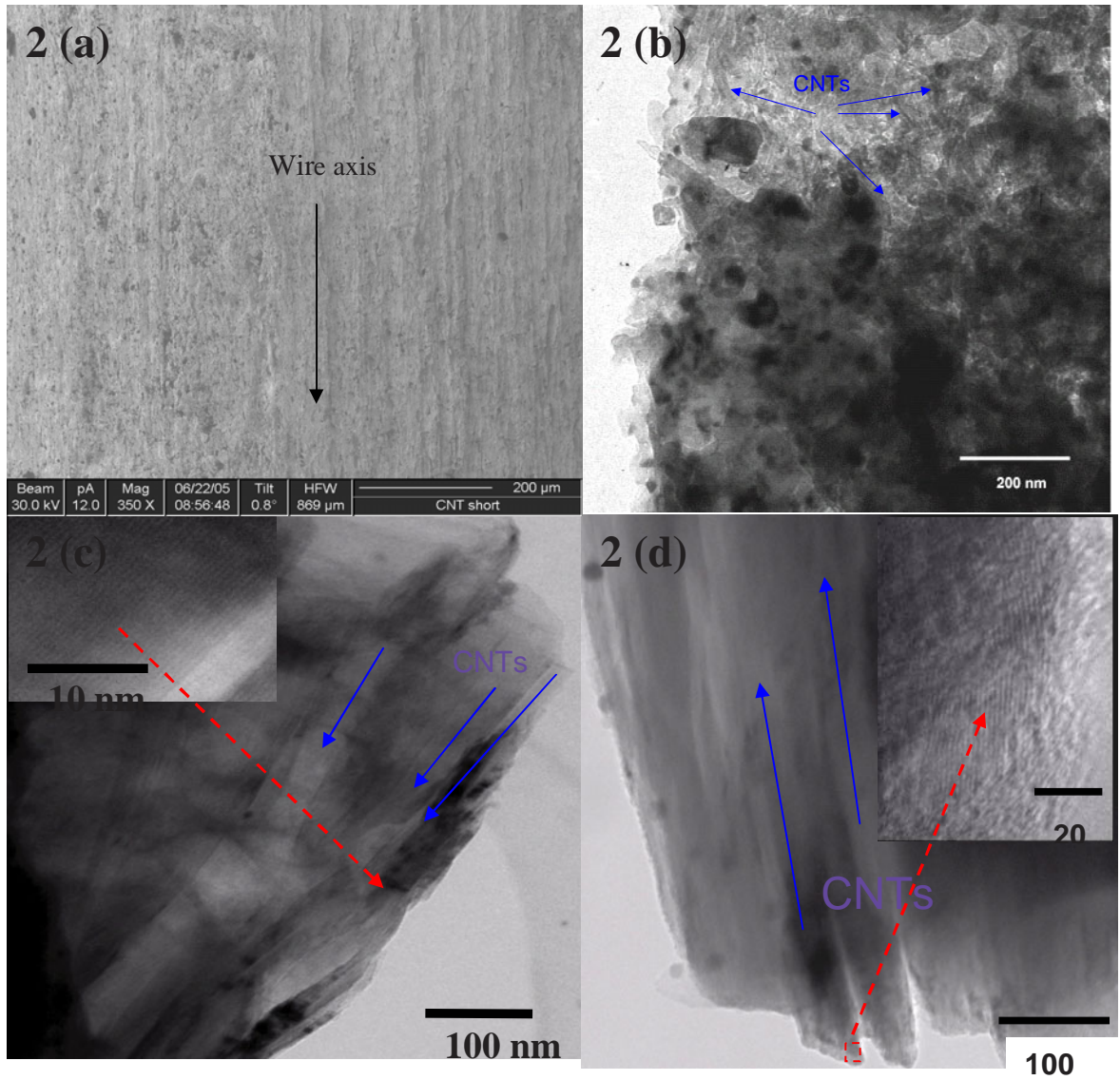

Fig. 2 (a) FIB-SEM micrographs of the CNT doped $MgB_2$ wire core, showing the elongated macrostructure along the wire axis, (b) Transmission electron micrographs (TEM) for the CNT doped $MgB_2$ pellet, showing the entangled CNTs randomly distributed in the $MgB_2$ matrix, (c) TEM image for the CNT doped $MgB_2$ wire, showing the a bundle CNTs in the same direction in the $MgB_2$ matrix. The inset in Fig 2 (c) is the high resolution image of CNT, and (d) TEM image for several parallel CNTs embedded in $MgB_2$. The inset in Fig 2 (d) is the high resolution lattice image of one of CNTs.

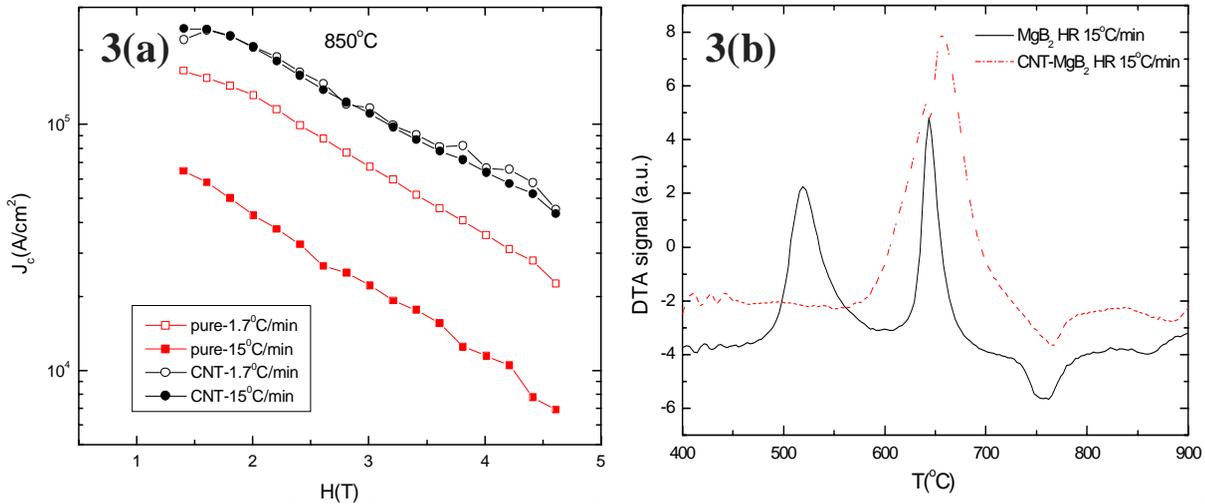

Fig. 3 (a) A comparison of $J_c(H)$ for the undoped and CNT doped $MgB_2$ wires processed with heating rate of 100°C/h and 900°C/h. There is no effect of heating rate on $J_c(H)$ for CNT doped $MgB_2$ wire but the fast heating rate severely degrades the $J_c(H)$. (b) The DTA curves for the undoped and CNT doped $MgB_2$ wires processed with heating rate of 15°C/min.

Furthermore, the CNTs have a strong effect on heat transfer during materials processing due to their high thermal conductivity. Fig. 3 (a) shows the heating rate effect on $J_c(H)$ behavior of the pure and CNT doped $MgB_2$ wires. We note that there is no effect of heating rate on $J_c(H)$ for the CNT doped $MgB_2$ wire while the $J_c(H)$ for pure $MgB_2$ wire at rapid heating rate of 15°C/min. is lower than that at slow heating rate of 1.7°C/min. by a factor of 3. In the rapid heating case for pure $MgB_2$ wire, the solid state reaction is inhomogeneous due to the large temperature gradient. On the other hand, the CNTs, due to their efficient heat transfer and acting as nucleation centers enhance the homogeneous reaction throughout the core. In the case of slow heating rate there is sufficient time to transfer the heat to the interior and hence allow the reaction to take place homogeneously in the entire core. This interpretation is further verified by the curves of differential thermal analysis (DTA) on the Fe sheathed pure $MgB_2$ and 10% CNT doped $MgB_2$ wires with heating rate at 15°C/min. as shown in Fig. 3 (b). There are two exothermic peaks for pure $MgB_2$ wire while for CNT doped $MgB_2$ wire these two peaks are merged to one broad peak. The first peak is due to the reaction between Mg and $B_2O_3$ while the second peak is attributable to the formation of $MgB_2$. The CNTs with high aspect ratio are desirable nucleation centers to trigger and propagate the $MgB_2$ formation reaction along the CNTs.

In summary, the alignment of CNTs in the CNT/$MgB_2$ composite wires can be achieved by readily scalable drawing technique. The aligned CNT doped $MgB_2$ wire shows an enhancement of magnetic $J_c$ by more than an order of magnitude in high field region. The aligned CNTs induce anisotropy in magnetic $J_c$ in relation to the direction of applied field and significantly improve heat transfer and dissipation during materials processing. In addition to the benefits of electrical and thermal conductivity of CNT doping the aligned CNTs in the wire axis direction will take full advantage of their unusual axial strength to enhance the mechanical properties such as tensile strength and flexibility. Studies on these properties are underway.

*Experimetal*

CNT doped $MgB_2$ wires were prepared using PIT method through a reaction in-situ process [25,26]. Powders of magnesium (99%) and amorphous boron (99%) were well mixed with 0 and 10 wt% of randomly distributed multi-wall carbon nano-tubes (OD: 20-30 nm and length: 0.5–2 µm) and thoroughly ground. Ultrasonic mixer was used to improve the homogeneity of mixed Mg, B and CNT. The Fe tube had an outside diameter (OD) of 10 mm, a wall thickness of 1 mm, and was 10 cm long with one end of the tube sealed. The mixed powder was filled into the tube and the remaining end was blocked using an aluminum bar. The composite was drawn to a 1 mm to 1.4 mm diameter wire through a series of more than 30 dies with reduction rate about 10% every drawing. For fabrication of multifilament wires, a bundle of single core wires were inserted to an iron tube and the composite was drawn to a 1 mm to 1.4 mm diameter wire.

Several short samples 2 cm in length were cut from the wire. These pieces were then sintered in a tube furnace at 800°C and 830°C for 30min with a heating rate of 3°C per minute, and finally furnace-cooled to room temperature. A high purity argon gas flow was maintained throughout the sintering process. An undoped sample was also made under the same conditions for the use as a reference sample. Sintering of wires at different heating rates was carried out at Cambridge. Two heating rates of 1.7 °C per minute and 15 °C per minute were used, followed by isothermal annealing at 850 °C for 30 min, and then furnace cooled to room temperature. The phase and crystal structure of all the samples was obtained from X-ray diffraction (XRD) patterns using a Philips (PW1730) diffractometer with Cu $K\alpha$ radiation. Differential thermal analysis (DTA) was performed to study the heating rate effect on $J_c$. The grain morphology and microstructure were also examined by scanning electron microscope (SEM) equipped with focused ion beam (FIB) and transmission electron microscope (TEM). The magnetization was measured at 5 and 20 K using a Physical Property Measurement System (PPMS, Quantum Design) in a time-varying magnetic field with sweep rate 50 Oe/s and amplitude 8.5T. Bar shaped samples with a diameter of 0.7 mm and length of 2.7 mm were cut from wire core for magnetic measurements. The magnetic measurements were performed by applying the magnetic field perpendicular and parallel to the wire sample axis. Since there is a large sample size effect on the magnetic $J_c$ for $MgB_2$ [27,28] all the samples for measurement were made to the same size for comparison. The magnetic $J_c$ was derived from the width of the magnetization loop using Bean's model. $J_c$ versus magnetic field has been measured up to 8.5 T.